\documentclass[11pt,twoside]{article}

\usepackage{pgf}
\usepackage[latin1]{inputenc}
\usepackage{amsfonts}
\usepackage{amssymb}
\usepackage{amsmath,multirow}
\usepackage{amsthm}
\usepackage{epsfig,array}
\usepackage{url}
\usepackage{hyperref}
\usepackage{fancyhdr}
\usepackage{footnote}
\usepackage{algorithm}
\usepackage[all]{xy}
\usepackage{siunitx}
\makeatletter
\renewenvironment{thebibliography}[1]
     {\section*{\refname}%
      \@mkboth{\MakeUppercase\refname}{\MakeUppercase\refname}%
      \list{\@biblabel{\@arabic\c@enumiv}}%
           {\settowidth\labelwidth{\@biblabel{#1}}%
            \leftmargin\labelwidth
            \advance\leftmargin\labelsep
            \@openbib@code
            \usecounter{enumiv}%
            \let\p@enumiv\@empty
            \itemsep=0pt
            \parsep=0pt
            \leftmargin=\parindent
            \itemindent=-\parindent
            \renewcommand\theenumiv{\@arabic\c@enumiv}}%

      \sloppy
      \clubpenalty4000
      \@clubpenalty \clubpenalty
      \widowpenalty4000%
      \sfcode`\.\@m}
     {\def\@noitemerr
       {\@latex@warning{Empty `thebibliography' environment}}%
      \endlist}
\makeatother

\date{}

\begin{document}

\centerline {\Large{\bf Self-Dual Skew Cyclic Codes over $\mathbb{F}_{q}+u\mathbb{F}_{q}$}}

\centerline{}

%% My definition
\newcommand{\mvec}[1]{\mbox{\bfseries\itshape #1}}

\centerline{
	\bf {Zineb Hebbache, Kenza Guenda, N. Tugba \"{O}zzaim, Mehmet \"{O}zen and T. Aaron Gulliver}}
\centerline{}

%\centerline{Faculty of Mathematics, USTHB}
%\centerline{
%USTHB, Laboratory of Algebra and number theory, BP 32 El Alia, Bab Ezzouar, Algeria}

\centerline{E-mail: {zinebhebache@gmail.com} (Z. Hebbache), {ken.guenda}@gmail.com (K. Guenda)\footnote{Faculty of Mathematics,
		USTHB, Laboratory of Algebra and Number Theory, BP 32 El Alia, Bab Ezzouar, Algeria}}

%\centerline{Department of Mathematics, Sakarya University, Sakarya, Turkey}
% \centerline{E-mail: E-mail addresses: ozen@sakarya.edu.tr (M. Özen)}

\centerline{}
%
%\centerline{Faculty of Mathematics, USTHB}
%\centerline{USTHB, Laboratory of Algebra and number theory, BP 32 El Alia, Bab Ezzouar, Algeria}

%\centerline{E-mail:{zinebhebache},{ken.guenda}@gmail.com\footnote{USTHB, Laboratory of Algebra and number theory, BP 32 El Alia, Bab Ezzouar, Algeria}}

%\centerline{Department of Mathematics, Sakarya University, Sakarya, Turkey}
\centerline{E-mail: ozen@sakarya.edu.tr (M. \"{O}zen), tugbaozzaim@gmail.com (N.T. \"{O}zzaim)\footnote{Department of Mathematics,
Sakarya University, Sakarya, Turkey}}

\centerline{E-mail: agullive@ece.uvic.ca (T. Aaron Gulliver)\footnote{Department of Electrical and Computer Engineering,
University of Victoria, PO Box 1700, STN CSC, Victoria, BC, Canada V8W 2Y2}}

\newtheorem{definition}{ Definition}
\newtheorem{prop}{ Proposition}
\newtheorem{cor}{ Corollary}
\newtheorem{lem}{ Lemma}
\newtheorem{rem}{ Remark}
\newtheorem{thm}{ Theorem}
\newtheorem{expl}{ Example}

{\bf Abstract}
{\emph{ In this paper, we give conditions for the existence of Hermitian self-dual $\Theta-$cyclic and $\Theta-$negacyclic codes
over the finite chain ring $\mathbb{F}_q+u\mathbb{F}_q$.
By defining a Gray map from $R=\mathbb{F}_q+u\mathbb{F}_q$ to $\mathbb{F}_{q}^{2}$,
we prove that the Gray images of skew cyclic codes of odd length $n$ over $R$ with even characteristic
are equivalent to skew quasi-twisted codes of length $2n$ over $\mathbb{F}_q$ of index $2$.
We also extend an algorithm of Boucher and Ulmer \cite{BF3} to construct self-dual skew cyclic codes
based on the least common left multiples of non-commutative polynomials over $\mathbb{F}_q+u\mathbb{F}_q$.
}}


\section{Introduction}
Codes over finite rings have been studied for many years and interest in these codes has increased recently due
to applications such as quantum and DNA systems by Nabil {\em et al.} \cite{Nbil}.
Boucher {\em et al.} \cite{ga} generalized the construction of linear codes via skew polynomial rings by using Galois rings instead of finite fields as coefficients.
Jitman {\em et al.} \cite{SSP} defined skew constacyclic codes over finite chain rings, in particular over the ring $\mathbb{F}_{p^{m}}+u\mathbb{F}_{p^{m}}$.
Recently, Ashraf {\em et al.} \cite{MG} studied the structural properties of skew cyclic codes over the finite semi-local ring $\mathbb{F}_{p^{m}}+v\mathbb{F}_{p^{m}}$ with $v^{2}=1$.
An advantage of skew cyclic codes over $\mathbb{F}_{p^{m}}+u\mathbb{F}_{p^{m}}$ with $u^{2}=0$ is that many
of the best linear codes over finite fields can be obtained using the Gray images of these codes.

Batoul {\em et al.} \cite{AKTN} gave conditions on the existence of self-dual codes derived from constacyclic codes over finite principal ideal rings.
Further, Boucher {\em et al.} (\cite{BD1}, \cite{BF3}) gave conditions on the existence of self-dual $\Theta-$cyclic and self-dual $\Theta-$negacyclic codes.
In this paper, we give necessary and sufficient conditions on the existence of self-dual $\Theta-$cyclic codes and $\Theta-$negacyclic codes over $\mathbb{F}_{q}+u\mathbb{F}_{q}$
where $\Theta$ is an automorphism of $\mathbb{F}_{q}+u\mathbb{F}_{q}$.
We also extend an algorithm of Boucher \cite{BF3} to construct self-dual skew cyclic codes based on the least common multiples of polynomials over the ring $\mathbb{F}_{p^{m}}+u\mathbb{F}_{p^{m}}$ with $u^{2}=0$.

The remainder of this article is organized as follows.
Some facts regarding Hermitian self-dual skew codes are recalled in Section 2.
In Section 3, we prove that the Gray images of skew cyclic codes of odd length $n$ over $R$ are equivalent to
skew quasi-twisted codes of length $2n$ over $\mathbb{F}_{q}$ with even characteristic using an approach by Amarra {\em et al.} \cite{AM}.
In Section 4, necessary and sufficient conditions are given on the existence of Hermitian self-dual skew codes generated by skew binomials and skew trinomials
when the characteristic of $R$ is odd.
Further, sufficient conditions on the existence of Hermitian self-dual skew cyclic codes of length $n$ over $R$ are given using the order of $q$ modulo $n$.
Section 5 gives an iterative construction of self-dual $\Theta-$cyclic codes over $R$ using the generator polynomials of self-dual $\Theta-$cyclic codes as
the least common left multiple of skew polynomials.

\section{Preliminaries}
This section provides some useful results regarding skew constacyclic codes over the ring $R=\mathbb{F}_q+u\mathbb{F}_q$.
First, we recall some facts concerning $R=\mathbb{F}_q+u\mathbb{F}_q$.
This is a commutative ring and can be defined as the quotient ring $R=\frac{\mathbb{F}_{q}[u]}{\langle u^{2}\rangle }$ with $u^2=0$.
This is an extension of degree $2$ of $\mathbb{F}_q$ so $|R|= q^2$.
This ring is principal with a unique maximal ideal $\langle u \rangle$.
To define skew polynomials over $R$, we first give the structure of the automorphism group of $R$ denoted by $Aut(R)$.
For $\theta \in \mathbb{F}_q$, $Aut(R)$ is given by
\[
Aut(R)=\Theta_{\theta, \beta};\, \theta \in Aut(\mathbb{F}_q) \text{ and } \beta \in \mathbb{F}_q^{\ast},
\]
where for $a+ub \in R$, $\Theta_{ \theta, \beta}(a+ub )= \theta(a)+ \beta \theta(b)u$ \cite[Corollary 2.1]{SSP}.
For simplicity, where no confusion arises, the subscripts $\theta$ and $\beta $ will be dropped.

Considering the finite chain ring $R$ and automorphism $\Theta $ of $R$ defined above, the set of formal polynomials
\begin{equation*}
\Re= R\lbrack x,\Theta ]=\left\{ a_{0}+a_{1}x+ \ldots +a_{n}x^{n}; a_{i}\in R
\text{ and }n\in \mathbb{N}^{\ast}\right\},
\end{equation*}
forms a ring under the usual addition of polynomials where multiplication is defined using the rule $ x \times a=\Theta (a)x$.
This multiplication is extended to all elements in $\Re$ by associativity and distributivity.
The ring $\Re$ is called a skew polynomial ring over $R$ and an element of $\Re$ is called a skew polynomial.
It is easily seen that the ring $\Re$ is non-commutative unless $\Theta$ is the identity automorphism of $R$.
According to \cite[Section 1]{ga}, $\Re$ is no longer left or right Euclidean, but left or right division can be defined for some elements.
From \cite[Section 2]{SSP}, the subring of the elements of $R$ that are fixed by $\Theta$ is $R^{\Theta}$.
\begin{prop}
\label{prop: bidon 31}
Let $R=\mathbb{F}_q+u\mathbb{F}_q$ with $u^{2}=0$,
$\Theta$ be an automorphism of $R$ with order $e$ and $\Re=R[x, \Theta]$.
Then the center $Z(\Re)$ of $\Re$ is $R^{\Theta}[x^{e}]$.
\end{prop}
\begin{proof}
For any integer $i\in N$, the power $x^{ei}$ is also in the center $Z(\Re)$ of $\Re$.
This follows from the fact that $e$ is the order of the automorphism $\Theta$ so we have that $x^{ei}(\alpha)=(\Theta^{e})^{i}(\alpha)x^{ei}=\alpha x^{ei}$
for any $\alpha \in R$.
This implies that $f=\alpha_{0}+\alpha_{1}x^{e}+\alpha_{2}x^{2e}+\dots +\alpha_{s}x^{se}$ with $\alpha_{i}\in R^{\Theta}$ a central element.
Conversely, for any $f\in Z(\Re)$ and $a\in R$, if $xf=fx$ and $af=fa,$ then $f\in R^{\Theta}[x^{e}]$.
\end{proof}

We restrict our study to the case where the length $n$ of codes is a multiple of the order of $\Theta$ and $\lambda$ is a unit in $R^{\Theta}.$
The following Proposition explains why the factors in the decomposition of the generator of a central monic polynomial into two monic polynomials always commute.
\begin{prop}
\label{lem:bidon 29}
Under the hypothesis of Proposition~\ref{prop: bidon 31}, if $f\in R^{\Theta}[x^{e}]$ is a monic polynomial which decomposes into a product of two
monic polynomials $h$ and $g$ as $hg$ over $\Re=R[x, \Theta]$, then $hg=gf$ in $\Re$.
\end{prop}
\begin{proof}
Since $hg$ is a central element we have $(hg)h = h(hg)$.
Therefore $h(gh-hg)=0$.
Since the leading coefficient of $h$ is invertible, $h$ is not a zero divisor, so that $hg=gh$ in $\Re$.
\end{proof}

A code of length $n$ over $R$ is a nonempty subset of $R^{n}$.
A code $C$ over $R$ is said to be linear if it is a submodule of the $R-$module $R^{n}$.
In this paper, all codes are assumed to be linear unless otherwise stated.

Given an automorphism $\Theta$ of $R$ and a unit $\lambda$ in $R$, a code $C$ is said to be skew constacyclic, or specifically, $\Theta-\lambda-$constacyclic,
if $C$ is closed under the $\Theta-\lambda-$constacyclic shift
\begin{equation*}
\rho_{\Theta, \lambda}: R^{n}\rightarrow R^{n},
\end{equation*}
defined by
\begin{equation*}
\rho_{\Theta, \lambda}((a_{0},a_{1}, \ldots,a_{n-1}))=(\Theta( \lambda a_{n-1}),\Theta(a_{0}),\dots,\Theta(a_{n-2})).
\end{equation*}
In particular, such codes are called skew cyclic and skew negacyclic when $\lambda$ is $1$ and $-1$, respectively.
When $\Theta$ is the identity automorphism, they become the classical constacyclic, cyclic and negacyclic codes.

When the order of $\Theta$ is $2$, the Hermitian dual code of a code $C$ of length $n$ over $R$ is defined using the Hermitian inner product
$\langle u, v\rangle _{H} = \sum_{i=0}^{n-1} u_{i}\Theta(v_{i})$ for $u=(u_{0}, u_{1},\dots,u_{n-1})$ and $v=(v_{0}, v_{1},\dots,v_{n-1})$ in $R^{n}$ as $ C^{\perp H }=\{v\in R^{n}|\left\langle v,c\right\rangle _{H}=0$ for all $c\in C\}$.
A code $C$ is said to be Hermitian self-dual if $C=C^{\perp H}.$

Given a right divisor $g(x)=\sum_{i=0}^{n-k-1} g_{i}x^{i}+x^{n-k}$ of $x^{n}-\lambda$,
a generator matrix of the $\Theta-\lambda-$constacyclic code $C$ generated by $g(x)$ is given by
\begin{equation}
\label{equa:matrix}
G=
\begin{bmatrix}
g_{0}& \dots & g_{n-k-1}& 1& 0& \dots& 0\\
0& \Theta(g_{0})&\dots & \Theta(g_{n-k-1})& 1& \dots& 0\\
0& \dots& \dots& \dots& \Theta^{2}(g_{n-k-1})& \dots& 0\\
\vdots& \ddots& \ddots& \ddots& \ddots& \ddots& \vdots\\
0& \dots& 0& \Theta^{k-1}(g_{0})& \dots& \Theta^{k-1}(g_{n-k-1}) & 1\\
\end{bmatrix}
\end{equation}
From \cite{SSP}, the ring automorphism $\phi$ on $\Re$ is defined as
\begin{equation}
{\label{eq:SDM1}}
\phi\left( \sum_{i=0}^{t} a_{i}x^{i}\right)=\sum_{i=0}^{t} \Theta(a_{i})x^{i}.
\end{equation}
According to \cite{SSP}, we have the following results which characterize the skew constacyclic and Hermitian self-dual skew constacyclic codes
when the order of $\Theta$ is $2$.
\begin{prop}(\cite{SSP})
\label{prop:bidon26}
Let $C$ be a code of even length $n$ over $R, \lambda^{2}=1$ and let $h, g \in \Re$ such that $g$ is a monic right divisor of $x^{n}-\lambda$ and $h(x)=\frac{x^{n}-\lambda}{g(x)}$.
Then the $\Theta-\lambda-$constacyclic code $C$ generated by $g(x)$ is a free $R-$module with $\vert C \arrowvert=\vert R\arrowvert^{n-deg(g(x))}$.
Furthermore, $C$ is a $\Theta-\lambda-$constacyclic code generated by $g(x)$ if and only if $C^{\perp H}$ is a $\Theta-\lambda-$constacyclic code generated by
\[
g^{\perp}=\phi (x^{deg(h(x))} \varphi(h(x))),
\]
where $\varphi: \Re \longrightarrow \Re S^{-1} $ is the ring anti-monomorphism defined by
\begin{equation}
\label{eq:SDM}
\varphi\left(\sum_{i=0}^{t} a_{i}x^{i}\right)=\sum_{i=0}^{t} x^{-i}a_{i},
\end{equation}
with $S=\{x^{i}; i\in \mathbb{N}\}$.
\end{prop}

\begin{prop}(\cite[Theorem 3.8]{SSP})
\label{prop: bidon28}
Assume that the order of $\Theta$ is $2$, $\lambda^{2}=1$ and $n$ is an even integer $n=2k$.
Let $g(x)=\sum_{i=0}^{k-1} g_{i}x^{i}+x^{k}$ be a right divisor of $x^{n}-\lambda$.
Then the $\Theta-\lambda-$constacyclic code generated by $g(x)$ is Hermitian self-dual if and only if
\begin{equation}
\label{eq:SDM3}
(\sum_{i=0}^{k-1} g_{i}x^{i}+x^{k})(\Theta^{-k-1}(g_{0}^{-1})+\sum_{i=1}^{k-1}\Theta^{i-k-1}(g_{0}^{-1}g_{k-i})x^{i}+x^{k})=x^{n}-\lambda.
\end{equation}
\end{prop}
This is called the self-dual skew condition.

\begin{definition}(\cite[Definition 2] {BF4})
\label{def:bidon4}
Let $R$ be a commutative finite ring.
The skew reciprocal polynomial of
\begin{equation*}
h=\sum_{i=0}^{m}h_{i}X^{i}\in \Re,
\end{equation*}
of degree $m$ is
\begin{equation*}
h^{\ast}=\sum_{i=0}^{m}X^{m-i}h_{i}=\sum_{i=0}^{m}\Theta^{i}\left( h_{m-i}\right) X^{i}.
\end{equation*}
The left monic skew reciprocal polynomial of $h$ is $h^{\natural}=\frac{1}{\Theta^{m}(h_{0})} h^{\ast}$.
\end{definition}

In the following, we focus on the relationship between skew $\lambda-$constacyclic codes, $\lambda-$constacyclic codes and $\lambda-$quasi-twisted codes over $R$.
\begin{prop}
Let $C$ be a skew $\lambda-$constacyclic code of length $n$ and let $\Theta$ be an automorphism of $R$ with order $e$.
If $\gcd(e,n)=1$, then $C$ is a constacyclic code of length $n$ over $R$.
\end{prop}
\begin{proof}
Let $C$ be a skew $\lambda-$constacyclicc code of length $n$ such that $\gcd(e,n)=1$ and let $\lambda$ be a unit of $R$ fixed by $\Theta$.
We know that there exist integers $ a_{1}, a_{2}$ such that
\begin{equation*}
a_{1}e+a_{2}n=1\Longrightarrow a_{1}e=1-a_{2}n.
\end{equation*}
We may assume that $a_{2}$ is a negative integer, so we can write $a_{1}e=1+\ell n$ where $\ell >0$.
Let $c(x)=c_{0}+c_{1}x+\dots+c_{n-1}x^{n-1}$ be a codeword in $C$.
Note that for
\[
x^{a_{1}e} c(x)= \Theta^{a_{1}e}(c_{0})x^{1+\ell n}+\Theta^{a_{1}e}(c_{1})x^{2+\ell n}+\dots +\Theta^{a_{1}e}(c_{n-1})x^{n+\ell n},
\]
over the ring $\Re /\langle x^{n}-\lambda \rangle$, we have $x^{n}=\lambda$ for any $a\in R$ and $\Theta^{e}(a)=a$.
This implies that
\[
x^{a_{1}e} C(x)= \lambda c_{n-1}+c_{0}x+ \dots +c_{n-2}x^{n-2} \in C.
\]
Thus, $C$ is a constacyclic code of length $n$.
\end{proof}

From \cite[Definition 2]{Gao}, we give the definition of $\Theta-\lambda-$quasi-twisted codes.
\begin{definition}
Let $C$ be a linear code of length $n$ over $R$ and $n=s \ell$.
Then $R$ is said to be a $\Theta-\lambda-$quasi-twisted code if for any
\begin{equation*}
	\left(\begin{array}{cc}
	c_{0,0},c_{0,1},\dots,c_{0,\ell-1},c_{1,0},c_{1,1},\dots, c_{1,\ell-1},\dots, \\
	c_{s-1,0},c_{s-1,1},\dots ,c_{s-1,\ell-1}
	\end{array}
	\right)\in C,
	\end{equation*}
	then
	\begin{equation*}
	\left(\begin{array}{cc}
	\lambda \Theta(c_{s-1,0}), \lambda \Theta(c_{s-1,1}), \dots, \lambda \Theta(c_{s-1,\ell-1}),\Theta(c_{0,0}), \Theta(c_{0,1}),\dots, \Theta(c_{0,\ell-1}), \dots, \\
	\Theta(c_{s-2,0}), \Theta(c_{s-2,1}), \dots, \Theta(c_{s-2,\ell-1})
	\end{array}
	\right)\in C.
	\end{equation*}
\end{definition}
If $\Theta$ is the identity map, we call $C$ a quasi-twisted code of index $\ell$ over $R$.

\begin{prop}
Let $C$ be a skew $\lambda-$constacyclic code of length $n=s\ell$ over $R$ and $\Theta$ be an automorphism with order $e$.
If $\gcd(e,n)=\ell$, then $C$ is a $\lambda-$quasi-twisted code of length $n$ with index $\ell$ over $R$.
\end{prop}

\begin{proof}
Let $n=s\ell$ and
\[
			\left(\begin{array}{cc}
			c_{0,0},c_{0,1},\dots,c_{0,\ell-1},c_{1,0},c_{1,1}, \dots, c_{1,\ell-1}, \dots, \\
			c_{s-1,0},c_{s-1,1},\dots,c_{s-1,\ell-1}
			\end{array}
			\right) \in C.
\]
Since $\gcd(e,n)=\ell,$ there exist integers $ a_{1}, a_{2}$ such that $ a_{1}e+a_{2}n=\ell$.
Therefore $a_{1}e=\ell-a_{2}n=\ell+jn$ where $j$ is a nonnegative integer.
Furthermore, let $\lambda$ be a unit of $R$ fixed by $\Theta$ and
			\begin{equation*}
			\Theta^{\ell+jn}\left(\begin{array}{cc}
			c_{0,0},c_{0,1}, \dots,c_{0,\ell-1},c_{1,0},c_{1,1}, \dots, c_{1,\ell-1}, \dots, \\
			c_{s-1,0},c_{s-1,1}, \dots,c_{s-1,\ell-1}
			\end{array}
			\right)
			\end{equation*}
			\begin{equation*}
			= \left(\begin{array}{cc}
			\Theta^{\ell+jn}(\lambda c_{s-1,0}),\Theta^{\ell+jn}(\lambda c_{s-1,1}), \dots,\Theta^{\ell+jn}(\lambda c_{s-1,\ell-1}),\Theta^{\ell+jn}(c_{0,0}),\dots,
			& \\  \Theta^{\ell+jn}(c_{0,\ell-1}), \dots, \Theta^{\ell+jn}( c_{s-2,0}),\Theta^{\ell+jn}(c_{s-2,1}), \dots,\Theta^{\ell+jn}(c_{s-2,\ell-1})
			\end{array}
			\right)
			\end{equation*}
Since the order of $\Theta$ is $e$, $\Theta^{\ell+jn}(a)=\Theta^{a_{1}e}(a)=a$ for any $a\in R$ which implies that
			\begin{equation*}
			\Theta^{\ell+jn}\left(\begin{array}{cc}
			c_{0,0},c_{0,1}, \dots,c_{0,\ell-1},c_{1,0},c_{1,1}, \dots, c_{1,\ell-1}, \dots, \\
			c_{s-1,0},c_{s-1,1}, \dots,c_{s-1,\ell-1} 	
			\end{array}
			\right)
			\end{equation*}
			\begin{equation*}
			= \left(\begin{array}{cc}
			\lambda c_{s-1,0}, \lambda c_{s-1,1}, \dots, \lambda c_{s-1,\ell-1},c_{0,0},\dots,
			&  \\  c_{0,\ell-1}, \dots,  c_{s-2,0},c_{s-2,1}, \dots,c_{s-2,\ell-1}
			\end{array}
			\right)\in C.
			\end{equation*}
Thus, $C$ is a $\lambda-$quasi-twisted code of length $n$ with index $\ell$ over $R$.
\end{proof}

\section{Gray Images of Skew Cyclic Codes with Odd Length}
In this section, we give a characterization of the Gray images of skew cyclic codes of odd length $n$ over $R$ with even characteristic, where $\Theta=\Theta_{\theta,1}$
is defined as in the previous section.
Let $\lambda= 1+u$ be a unit of $R$.
Then we have that $\lambda^{n}=1$ if $n$ is even and $\lambda^{n}=\lambda$ if $n$ is odd.
Thus, we only consider the properties of skew $\lambda-$constacyclic codes of odd length in this section.

We know that every element of $R$ can be expressed as $a+ub$ where $a,b \in \mathbb{F}_q$.
According to Ling {\em et al.} \cite{SP}, we have the following Gray map
\[
\Phi: \; R^{n} \rightarrow \mathbb{F}_{q}^{2n}
\]
\begin{equation}
\label{equa:map}
\Phi(a+ub)=(b, a+b),
\end{equation}
where $ a, b \in \mathbb{F}_q^{n}$.
Following \cite{SW}, the Lee weight is defined as the Hamming weight of the Gray image
\[
w_{L}(a+ub)=w_{H}(b)+w_{H}(a+b), \mbox{ for } a,b\in \mathbb{F}_{q}^{n}.
\]
The Lee distance of $x,y\in R^{n}$ is defined as $w_{L}(x-y)$.
Thus, the Gray map $\Phi$ is a linear isometry from ($R^{n}$, Lee distance) to ($\mathbb{F}_q^{2n}$, Hamming distance).

Let $\sigma: \mathbb{F}_q^{2n} \rightarrow \mathbb{F}_q^{2n}$ be the skew quasi-twisted shift operator defined by
\[
\sigma (a^{(0)}\mid a^{(1)})=(\rho_{\Theta, \lambda}(a^{(0)}) \mid \rho_{\Theta, \lambda} (a^{(1)})),
\]
where $a^{(0)}, a^{(1)} \in \mathbb{F}_q^{n},
\mid$ is vector concatenation and $\rho_{\Theta, \lambda}$ is the skew constacyclic shift operator as defined in the previous section.
A linear code $C$ of length $2n$ over $\mathbb{F}_{q}$ is said to be skew quasi-twisted of index $2$ if $\sigma (C)=C$.

\begin{prop}
\label{prop:bidon5}
With the previous notation, we have $\Phi \circ \rho_{\lambda, \Theta} = \sigma \circ \Phi$.
 \end{prop}
\begin{proof}
Let $r=(r_{0}, r_{1}, \dots, r_{n-1})\in R^{n}$ where $r_{i}=a_{i}+ub_{i}, 0\leq i \leq n-1$.
Then we can write $\Phi(r)=(b_{i}, a_{i}+ub_{i})$, so that
\begin{center}
$\Phi (\rho_{\Theta, \lambda}(r))=
(\Theta(\lambda b_{n-1}),\Theta(b_{0}),\dots,\Theta(b_{n-2}), \Theta(\lambda a_{n-1})+ \Theta(\lambda b_{n-1})$,\\
$\Theta(a_{0})+\Theta (b_{0}),\dots, \Theta(a_{n-2})+\Theta(b_{n-2}))$.
\end{center}
On the other hand
\[
\begin{array}{ccl}
\sigma(\Phi(r))&=& \sigma (b_{0}, b_{1}, \dots, b_{n-1}, a_{0}+b_{0}, a_{1}+b_{1}, \dots, a_{n-1}+b_{n-1})\\
&=&\left(\begin{array}{cc}
\Theta(\lambda b_{n-1}),\Theta(b_{0}),\dots,\Theta(b_{n-2}), \Theta(\lambda a_{n-1})+ \Theta(\lambda b_{n-1}),
& \\ \Theta(a_{0})+\Theta (b_{0}),\dots, \Theta(a_{n-2})+\Theta(b_{n-2}),
\end{array}
\right),
\end{array}
\]
and the result follows.
\end{proof}

\begin{thm}
\label{thm: bidon28}
Let $C$ be a code of length $n$ over $R$ and $\Theta:=\Theta_{\theta,1}$ where $\theta$ is an automorphism of $\mathbb{F}_{q}$.
Then $C$ is a skew $\lambda-$constacyclic code of length $n$ over $R$ if and only if $\Phi (C)$
is a skew quasi-twisted code of length $2n$ over $\mathbb{F}_{q}$ of index $2$.
\end{thm}
\begin{proof}
If $C$ is a $\Theta-\lambda-$constacyclic code, then $\rho_{\Theta, \lambda}(C)=C$.
We have $\Phi \left(\rho_{\Theta, \lambda}(C)\right)=\Phi\left(C\right)$, and from Proposition~\ref{prop:bidon5}
\begin{equation*}
\sigma \left(\Phi(C)\right) =\Phi \left( \rho_{\Theta, \lambda}(C) \right) =\Phi(C).
\end{equation*}
Hence, $\Phi(C)$ is a skew quasi-twisted code of index $2$.
Conversely, if $\Phi(C)$ is a skew quasi-twisted code of index $2$, then
\[
\sigma \left(\Phi(C)\right) =\Phi(C).
\]
Proposition~\ref{prop:bidon5} gives that
\begin{equation*}
\Phi(C)= \sigma \left(\Phi(C)\right) =\Phi \left( \rho_{\Theta, \lambda}(C) \right),
   \end{equation*}
and since $\Phi$ is injective, it follows that $\rho_{\Theta, \lambda}(C)= C$.
\end{proof}

\begin{thm}
	Define $\upsilon : \Re/ \langle x^{n}-1\rangle \rightarrow \Re/ \langle x^{n}-\lambda\rangle$ as
	\[
	\upsilon(c(x))= c(\lambda x).
	\]
	If $n$ is odd, then $\upsilon$ is a left $R-$module isomorphism.
\end{thm}
\begin{proof}
	The proof is straightforward starting from the fact that if $n$ is odd
	\[
	c(x) \equiv b(x) \bmod (x^{n}-1),
	\]
	if and only if $c(\lambda x) \equiv b(\lambda x) \bmod (x^{n}-\lambda)$.
\end{proof}

Immediate consequences of this theorem are given as follow
\begin{cor}
	$I$ is an ideal of $\Re / \langle x^{n}-1\rangle $ if and only if $\upsilon(I)$ is an ideal of $\Re/ \langle x^{n}-\lambda\rangle $.
\end{cor}
\begin{cor}
\label{prop: bidon 20}
Let $\mu$ be the permutation of $R^{n}$ with $n$ odd such  $ \mu(c_{0}, c_{1}, \dots, c_{n-1})=(c_{0}, (1+u)c_{1}, \dots, (1+u)^{n-1} c_{n-1})$
and $C$ be a non-empty subset of $R^{n}$.
Then $C$ is a skew cyclic code of length $n$ if and only if $\mu(C)$ is a skew $\lambda-$constacyclic code of length $n$ over $R$.
\end{cor}

We introduce the following permutation of $\mathbb{F}_{q}^{2n}$ from \cite{QZ} which is called the Nechaev permutation.
This will be useful in studying cyclic codes over $\mathbb{F}_{q}$.
\begin{definition}
Let $n$ be an odd integer and $\psi$ be the permutation of $\{0, 1, 2, \dots, 2n-1\}$ given by
	\begin{equation*}
	\psi=(1, n+1)(3, n+3)\dots (2i+1, n+2i+1) \dots (n-2, 2n-2).
	\end{equation*}
The Nechaev permutation is the permutation $\varrho$ of $\mathbb{F}_{q}^{2n}$ defined by
	\begin{equation*}
	\varrho(r_{0}, r_{1}, \dots, r_{2n-1})=(r_{\psi(0)}, r_{\psi(1)}, \dots, r_{\psi(2n-1)}).
	\end{equation*}
\end{definition}

\begin{prop}
$\Phi \circ \mu=\varrho \circ \Phi$.
\end{prop}
\begin{proof}
Let $(r_{0}, r_{1}, \dots, r_{n-1}) \in R^{n}$ where $r_{i}=a_{i}+ub_{i}, 0\leq i \leq n-1.$ From
\[
\mu(r)=(r_{0}, (1+u)r_{1}, \dots, (1+u)^{n-1} r_{n-1}),
\]
It follow that
\[
\begin{array}{ccl}
(\Phi\mu)(r)&=&(b_{0}, a_{1}+b_{1}, b_{2}, \dots, a_{n-2}+b_{n-2}, b_{n-1}, a_{0}+b_{0}, b_{1}, a_{2}+b_{2},\\
&&\hspace*{0.3in} \dots,  b_{n-2}, a_{n-1}+b_{n-1}).
\end{array}
\]
On the other hand, since
\[
\Phi(r)=(b_{0}, b_{1}, \dots, b_{n-1}, a_{0}+b_{0}, a_{1}+b_{1}, \dots, a_{n-1}+b_{n-1}),
\]
\[
\begin{array}{ccl}
(\varrho\Phi)(r)&=&\varrho(b_{0}, b_{1}, \dots, b_{n-1}, a_{0}+b_{0}, a_{1}+b_{1}, \dots, a_{n-1}+b_{n-1})\\
&=&(b_{0}, a_{1}+b_{1}, b_{2}, \dots, a_{n-2}+b_{n-2}, b_{n-1}, a_{0}+b_{0}, b_{1}, a_{2}+b_{2},\\
&&\hspace*{0.3in} \dots,  b_{n-2}, a_{n-1}+b_{n-1}).
\end{array}
\]
Therefore $(\Phi\mu)(r)= (\varrho\Phi)(r)$ and so $\Phi \mu=\varrho \Phi$.
\end{proof}
\begin{cor}
The Gray image of a skew cyclic code of length $n$ over $R$ is equivalent to a skew quasi-twisted code of length $2n$  over $\mathbb{F}_{q}$ of index $2$.
\end{cor}
\begin{proof}
From Proposition \ref{prop: bidon 20}, a code $C$ of length $n$ over $R$ is skew cyclic if and only if $\mu(C)$ is a skew $\lambda-$constacyclic code.
By Theorem \ref{thm: bidon28}, this is true if and only if $\Phi(\mu(C))$ is a skew quasi-twisted code of index $2$ over $\mathbb{F}_q$,
i.e. if and only if $\varrho(\Phi(C))$ is a skew quasi-twisted code of index $2$ over $\mathbb{F}_q$.
\end{proof}

\section{Hermitian Self-Dual Skew Codes over $\mathbb{F}_q+u\mathbb{F}_q$}
In this section, we give necessary and sufficient conditions for the existence of Hermitian self-dual $\Theta-$cyclic and self-dual $\Theta-$negacyclic codes.
\begin{prop}
\label{prop:bidon8}
Let $R=\mathbb{F}_q+u\mathbb{F}_q$ be a finite ring with $q=p^{m}$, $p$ odd prime number of the residue field $\mathbb{F}_{q}$ and $\Theta \in \text{Aut}(R)$.
Assume that the order of $\Theta$ is $2$, $\lambda \in R $ such that $\lambda^{2}=1$ and $n$ is even, denoted by $n=2k$.
\begin{enumerate}
\item If $k$ is even, then there exist Hermitian self-dual $\Theta-$cyclic codes over $R$ if and only if $p^{m}\equiv 1 \bmod 4$.
\item If $k$ is odd, then there exist Hermitian self-dual $\Theta-$negacyclic codes over $R$.
\end{enumerate}
\end{prop}
\begin{proof}
Assume that the order of $ \Theta$ is $2$, $\lambda^{2}=1$ and $n$ is even denoted by $n=2k$.
Let $g(x)=\sum_{i=0}^{k-1}g_{i}x^{i}+x^{k}$ be a right divisor of $x^{n}-\lambda$.
Now suppose that there is a Hermitian self-dual $\Theta-\lambda-$constacyclic code generated by $g(x)$.
Then by Proposition~\ref{prop: bidon28}, $g(x)$ satisfies (\ref{eq:SDM3}), so $-\lambda=g_{0}\Theta^{-k-1}(g_{0}^{-1})$.
Since $\lambda$ is fixed by $\Theta$, it follows that $\lambda=-\Theta^{k+1}(g_{0})g_{0}^{-1}$.
As the order of $\Theta$ is $2$, we have the following.
\begin{enumerate}
\item If $k$ is even then $\lambda=-\Theta(g_{0})g_{0}^{-1}=\frac{-\Theta(g_{0})}{g_{0}}=\frac{\frac{-1}{g_{0}}}{g_{0}}=\frac{-1}{g_{0}^{2}}$
and from \cite[Lemma 4.2]{SJH} we have that $-1$ is a square in $R$ if and only if $p^{m}\equiv 1 \bmod 4$.
Then there exists $g_{0}\in R$ such that $g_{0}^{2}=-1$, so $\lambda=\frac{-\Theta(g_{0})}{g_{0}}=\frac{-1}{g_{0}^{2}}=1$.
This implies that there exists a Hermitian self-dual $\Theta-$cyclic code over $R$ if and only if $p^{m}\equiv 1 \bmod 4$ and $k$ is even.
\item If $k$ is odd then $\lambda=-\Theta^{k+1}(g_{0})g_{0}^{-1}=-g_{0}g_{0}^{-1}=-1,$ so there exists a Hermitian self-dual $\Theta-$negacyclic code over $R$.
\end{enumerate}
\end{proof}

\begin{prop}
\label{cor:bidon8}
	Let $R=\mathbb{F}_q+u\mathbb{F}_q$ be a finite ring with $q=p^{m}$, $p$ prime number of the residue field $\mathbb{F}_{q}$ and $\Theta \in \text{Aut}(R)$.
	Assume that the order of $\Theta$ is $2$, $\lambda \in R $ such that $\lambda^{2}=1$ and $n$ is even denoted by $n=2k$. Let $g(x)=\sum_{i=0}^{k-1}g_{i}x^{i}+x^{k}$ be a right divisor of $x^{n}-\lambda$ with $g_{0}$ is a unit in $R$.
If $p$ is even, then there exist Hermitian self-dual $\Theta-$cyclic codes over $R$ for any integer $k$.
\end{prop}
\begin{proof}
	Suppose that there is a Hermitian self-dual $\Theta-\lambda-$constacyclic code generated by $g(x)$. Then by Proposition~\ref{prop: bidon28}, $g(x)$ satisfies (\ref{eq:SDM3}), so $-\lambda=g_{0}\Theta^{-k-1}(g_{0}^{-1})$. Since $\lambda$ is fixed by $\Theta$, it follow that $\lambda=-\Theta^{k+1}(g_{0})g_{0}^{-1}$.
Further, if $p$ has even characteristic then $\lambda=\Theta^{k+1}(g_{0})g_{0}^{-1}$.
As the order of $\Theta$ is $2$ we have the following.
	\begin{enumerate}
		\item If $k$ is even then $\lambda=\Theta(g_{0})g_{0}^{-1}=\frac{1}{g_{0}}g_{0}^{-1}=\frac{1}{g_{0}^{2}}$ as $g_{0}$ is a unit in $R$,
so $\lambda=\frac{1}{g_{0}^{2}}=1$.
This implies that there exists a Hermitian self-dual $\Theta-$cyclic code over $R$.
		\item If $k$ is odd then $\lambda=\Theta^{k+1}(g_{0})g_{0}^{-1}=g_{0}g_{0}^{-1}=1$, so there exists a Hermitian self-dual $\Theta-$cyclic code over $R$.
	\end{enumerate}
\end{proof}

We now consider the existence of Hermitian self-dual $\Theta-$cyclic and self-dual $\Theta-$negacyclic codes over $R$ of odd characteristic generated by
skew binomials and skew trinomials.
\begin{prop}
\label{prop:bidon4}
Let $R=\mathbb{F}_q+u\mathbb{F}_q$ be a finite ring with $q=p^{m}$, $p$ an odd prime number and $q\equiv 1 \bmod 4$. Consider $r\in \mathbb{N}$ and $\Theta \in \text{Aut}(R)$.
Assume that the order of $\Theta$ is $2$ and $\theta \in \text{Aut}(\mathbb{F}_{q})$ defined by $\theta :x\mapsto x^{p^{r}}$ and $n$ is even denoted by $n=2k$.
\begin{enumerate}
\item There are no Hermitian self-dual $\Theta-$cyclic or self-dual $\Theta-$negacyclic codes over $R$ generated by a skew trinomial.
\item There are Hermitian self-dual $\Theta-$cyclic and self-dual $\Theta-$negacyclic codes over $R$ generated by a skew binomial under the following conditions.
\begin{enumerate}
\item[(i)]There exists a Hermitian self-dual $\Theta-$cyclic code over $R$ if and only if $p\equiv 3 \bmod 4$, $k$ and $m$ are even and $r$ is odd.
\item[(ii)]There exists a Hermitian self-dual $\Theta-$negacyclic code over $R$ if and only if $p\equiv 1 \bmod 4$ or $p\equiv 3 \bmod 4$, $k$ is odd and $m$ and $r$ are even.
\end{enumerate}
\end{enumerate}
\end{prop}

\begin{proof}
Consider $g=g_{0}+g_{j}x^{j}+x^{k} \in \Re$ and let $\Theta= \Theta_{\theta, 1}$ be an automorphism of $R$ as defined in Section 2.
The $(\Theta , \lambda)-$constacyclic code generated by $g(x)$ is Hermitian self-dual if and only if $g$ satisfies (\ref{eq:SDM3}), i.e.
\begin{equation*}
(g_{0}+g_{j}x^{j}+x^{k})(\Theta^{-k-1}(g_{0}^{-1})+\Theta^{j-k-1}(g_{0}^{-1}g_{k-j}) x^{j}+x^{k}) =x^{n}-\lambda,
\end{equation*}
and from Proposition~\ref{prop:bidon8} we have
\begin{displaymath}
\lambda=
\left\{ \begin{array}{ll}
 1  & \textrm{if} $ k $ \textrm{is even};   \\
-1  & \textrm{if} $ k $  \textrm{is odd}.
\end{array}\right.
\end{displaymath}
Considering this skew polynomial relation, one obtains the equivalent conditions
\[
\begin{array}{l}
g_{0}\Theta^{j-k-1}(g_{0}^{-1}g_{k-j})+g_{j}\Theta^{j-k-1}(g_{0}^{-1})=g_{k-j}+g_{j}g_{0}^{-1}=0,\\
g_{j}\Theta^{2j-k-1}(g_{0}^{-1}g_{k-j})=g_{j} g_{0}^{-1}g_{k-j}=0,\\
g_{0}+\Theta^{-1}(g_{0}^{-1})=g_{0}^{2}+1=0, \mbox{ and}\\
g_{j} +\Theta^{j-1}(g_{0}^{-1} g_{k-j})=g_{j}+g_{0} g_{k-j}^{-1}=0.\\
\end{array}
\]
If $g_{j} \neq 0$, this system of equations has no solution, so there is no Hermitian self-dual $\Theta-$code over $R$ generated by a skew trinomial.
For part 2, if $g_{j}=0$, $g(x)=g_{0}+x^{k} \in \Re$, then the skew binomial $g$ is a skew reciprocal polynomial
that generates a self-dual skew $\lambda-$constacyclic code
if and only if $g$ satisfies (\ref{eq:SDM3}), i.e.
\begin{equation*}
(g_{0}+x^{k})(\Theta^{-k-1}(g_{0}^{-1})+x^{k})=x^{n}-\lambda, \text{ with } \lambda\in \{-1, 1\}.
\end{equation*}
This skew polynomial relation gives the equivalent condition
\[
g_{0}+\frac{1}{\Theta(g_{0}^{-1})}=g_{0}+\lambda g_{0}^{-1}=g_{0}+\lambda\Theta(g_{0})=g_{0}^{2}+1=0.
\]

Let $g_{0}=a_{0}+u b_{0}\in R$ with $a_{0}, b_{0} \in \mathbb{F}_{q}$.
If $g_{0}^{2}=-1$, we have $(a_{0}+u b_{0})(a_{0}+u b_{0})=a_{0}^{2}+2ua_{0}b_{0}=-1$ which implies that $a_{0}^{2}=-1$ and $2ua_{0}b_{0}=0$
so $b_{0}=0$ and then $\Theta(g_{0})=\theta(g_{0})$.
We have the following two cases.
\begin{enumerate}
\item There exists $g_{0} \in R$ such that $g_{0}+\Theta(g_{0})=g_{0}^{2}+1=0 $ if and only if $ p\equiv 3\bmod 4$, $k$ and $m$ are even, and $r$ is odd.
Suppose that $p\equiv 3 \bmod 4$, $m \equiv 0 \bmod 2$ and $r\equiv 1 \bmod 2$.
Then $-1$ is a square in $R$ and one can consider $g_{0} \in R$ such that $g_{0}^{2}=-1$.
As $r$ is odd, $p^{r}\equiv 3 \bmod 4$ so $ p^{r}-1\equiv 2 \bmod 4$ and $\frac{p^{r}-1}{2}\equiv 1 \bmod 2$.
Therefore
\[
g_{0}^{p^{r}-1}=(g_{0}^{2})^{\frac{p^{r}-1}{2}}=(-1)^{^{\frac{p^{r}-1}{2}}}=-1,\mbox{ i.e } g_{0}+\theta(g_{0})=g_{0}+\Theta(g_{0})=0.
\]
Conversely, consider $g_{0}$ in $R$ such that $g_{0}+\Theta(g_{0})=g_{0}^{2}+1=0$ and suppose that $ p\equiv 1\bmod 4$.
Then $-1$ is a square in $\mathbb{F}_{p} $ so $g_{0}$ belongs to $\mathbb{F}_{p}$ and $g_{0}$ is left fixed by $\theta$.
The equality $g_{0}+\Theta(g_{0})=g_{0}+\theta(g_{0})=0 $ implies that $2g_{0}=0$, which is impossible as $p$ is odd.
Therefore $p\equiv 3 \bmod 4$ and as $-1$ is a square in $R$, $m$ must be even.
Then $g_{0}+\Theta(g_{0})=g_{0}^{2}+1=0$ which gives
\[
-1=\frac{\Theta(g_{0})}{g_{0}}=\frac{\theta(g_{0})}{g_{0}}=g_{0}^{p^{r}-1}=( g_{0}^{2})^{\frac{p^{r}-1}{2}},
\]
so $\frac{p^{r}-1}{2}$ is odd and $p^{r}-1\equiv 2 \bmod 4$.
As $p\equiv 3 \bmod 4$, $r$ must be odd.

\item There exists $g_{0} $ in $R$ such that $g_{0}-\Theta(g_{0})=g_{0}^{2}+1=0$ if and only if $p\equiv 1 \bmod 4$ or $p\equiv 3 \bmod 4$,
$k$ is odd, and $m$ and $r$ are even.
If $p\equiv 1 \bmod 4$ then $g_{0} \in \mathbb{F}_{p} $ is left fixed by $\theta$.
Thus, $g_{0}-\Theta(g_{0})=g_{0}-\theta(g_{0})=0$.
If $p\equiv 3 \bmod 4$ and $r\times m\equiv 0 \bmod 2$, then $-1$ has a square root in $R$ and $p^{r}-1\equiv 0 \bmod 4$, so $\frac{p^{r}-1}{2}\equiv 0 \bmod 2$.
Consider $g_{0} \in R$ such that $g_{0}^{2}=-1$ so then
\[
g_{0}^{p^{r}-1}=\left( g_{0}^{2}\right) ^{\frac{p^{r}-1}{2}}=1,\mbox{ i.e. } g_{0}-\theta(g_{0})=g_{0}-\Theta(g_{0})=0.
\]
Conversely, consider $g_{0}\in R$ such that $g_{0}-\Theta(g_{0})=g_{0}^{2}+1=0$.
Therefore $-1$ is a square in $R$ and either $p\equiv 1\bmod 4$ or $p\equiv 3 \bmod 4$ and $m\equiv 0 \bmod 2$.
If $p\equiv 3 \bmod 4$ and $r\equiv 1\bmod 2$ then $p^{r}-1\equiv 2 \bmod 4$, so $\frac{p^{r}-1}{2}\equiv 1 \bmod 2$.
Thus $g_{0}^{p^{r}-1}=-1$, which contradicts the hypothesis $g_{0}-\Theta(g_{0})=g_{0}-\theta(g_{0})=0$
because $R$ has odd characteristic.
Therefore, $p\equiv 1 \bmod 4$ or $p\equiv 3 \bmod 4$ and $r\times m\equiv 0 \bmod 2$.
\end{enumerate}
\end{proof}

In the following theorem, we give sufficient conditions for the existence of Hermitian self-dual skew cyclic codes of length $n$ over $R$
based on the order of $q$ modulo $n$ where $q$ is odd.
Let $n$ be a positive integer, and
denote by $ord_{n}(p^{m})$ the multiplicative order of $p^{m}$ modulo $n$ which is the smallest integer $j$ such that $(p^{m})^{j}\equiv 1 \bmod n$.
Further, let $\mbox{lcm} (a, b)$ be the least common multiple of $a$ and $b$.

\begin{thm}
Let $R=\mathbb{F}_q+u\mathbb{F}_q$ with $q= p^{m}$ odd,
$\Theta \in \text{Aut}(R)$ and $n$ be an even integer denoted by $n=2k$ with $k$ odd.
If $ord_{n}(p^{m})$ is even, then there are no non-trivial $\Theta-$cyclic Hermitian self-dual codes of length $n$ over $R$.
\end{thm}
\begin{proof}
Let $n$ be an even integer denoted by $n=2k$ with $(2, k)=1$ and $(2, p^{m})=(k, p^{m})=1$ such that $k$ and $p^{m}$ are odd.
Then $ord_{n}(p^{m})=\mbox{lcm}(ord_{2}(p^{m}), ord_{k}(p^{m}))$ is even, and thus $ ord_{2}(p^{m})$ or $ ord_{k}(p^{m})$ must be even.
Suppose that $ord_{k}(p^{m})$ is even.
Then there exist $1\leq i \leq ord_{k}(p^{m})$ such that $(p^{m})^{i}\equiv -1 \bmod k$, and so
$q^{i}(n-2)\equiv-(n-2) \bmod n$ with $i \leq ord_{k}(p^{m})$.
Thus, there is at least one class that is reversible, $C_{n-2}$.
Then from \cite[Theorem 4.5]{AKTN}, there are no non-trivial $\Theta-$cyclic Hermitian self-dual codes.
\end{proof}

\section{Construction of Self-Dual $\Theta-$Cyclic Codes over $R$}
%\subsection{Construction Algorithm of Self-Dual $\Theta-$Cyclic Codes}
Before giving the construction of self-dual $\Theta-$cyclic codes,
we provide some results which will be useful later.
The following lemma is easy to prove as $\mathbb{F}_q$ can be considered as a sub-ring of $R$.
This result is similar to \cite[Remark 3.2]{KG}.
\begin{lem}
\label{lem:bidon 9}
	Let $R=\mathbb{F}_q+u\mathbb{F}_q$ with $u^{2}=0$ and residue field $\mathbb{F}_q$ of characteristic $p$.
	Consider $s, t\in \mathbb{N}$ and let $k=t\times p^{s}$ be an integer such that $\gcd(t,p)=1$.
	Then the factorization of $x^{n}-1$ into polynomials over $R=\mathbb{F}_q+u\mathbb{F}_q$ is the same as the factorization over $\mathbb{F}_q$.
\end{lem}

Recall that the center of $\mathbb{F}_q[x, \theta]$ is the commutative polynomial ring
$Z(\mathbb{F}_q[x, \theta] )=(\mathbb{F}_q)^{\theta}\left[x^{\lvert \theta \rvert}\right]$ where $(\mathbb{F}_q)^{\theta}$ is
the fixed field of $\theta$ and $\lvert \theta \rvert$ is the order of $\theta$.
Using Lemma~\ref {lem:bidon 9}, we prove the following proposition.
\begin{prop}
\label{prop:bidon 9}
	Let $\Theta \in \text{Aut}(R)$ and $\Re=R[x; \Theta]$, and assume that the order of $\theta$ is $2$.
	Consider $s, t\in \mathbb{N}$ and let $n=2k=2\times t\times p^{s}$ be an integer such that $\gcd(t,p)=1$.
	Then $x^{n}-1$ has the following decomposition in $\Re$
	\begin{equation}
	\label{eq:SDM4}
	x^{n}-1=f_{1}^{p^{s}}(x^{2}) \dots f_{l}^{p^{s}}(x^{2}),
	\end{equation}
	where the $f_{i}(x^{2})= f_{i}^{\natural}(x^{2})$ are pairwise coprime polynomials in
	$(\mathbb{F}_q)^{\theta}\left[x^{2}\right]=(\mathbb{F}_q)^{\theta}\left[y\right]$ which are divisors of $y^{t}-1$.
\end{prop}
\begin{proof}
	Let $\Theta \in \text{Aut}(R)$ and $\theta \in $ Aut$ \left(\mathbb{F}_{q} \right)$ of order $2$.
	Consider $s, t\in \mathbb{N}$ and let $k=t\times p^{s}$ be an integer. Then, from \cite[Lemma 27]{BF3}, we have that
	\[
	y^{t}-1\in (\mathbb{F}_q)^{\theta}\left[y\right]=(\mathbb{F}_q)^{\theta}\left[x^{2}\right] \subset \mathbb{F}_q[x, \theta],
	\]
	factors in $(\mathbb{F}_q)^{\theta}\left[y\right]$ as a product $f_{1}(y)\dots f_{l}(y)$ of pairwise coprime polynomials of minimal degree
	such that $f_{i}(y)= f_{i}^{\natural}(y)$.
	Furthermore, we have that
	\begin{equation*}
	\left((x^{2})^{t}-1\right)^{p^{s}}=\left(y^{t}-1\right)^{p^{s}}=y^{t{p^{s}}}-1,
	\end{equation*}
	and according to Lemma~\ref {lem:bidon 9}, if $\gcd(t, p)=1$ then
	\begin{equation}
	\label{eq:SDM5}
	x^{n}-1=\left(y^{t}-1\right)^{p^{s}}=f_{1}^{p^{s}}(y) \dots f_{l}^{p^{s}}(y),
	\end{equation}
	where the $f_{i}(y)= f_{i}^{\natural}(y)$ are pairwise coprime polynomials in
	$(\mathbb{F}_q)^{\theta}\left[x^{2}\right]=(\mathbb{F}_q)^{\theta}\left[y\right]$ which are divisors of $y^{t}-1$.
\end{proof}

We require the following proposition.
\begin{prop}
\label{prop: bidon 7}
Let $R=\mathbb{F}_q+u\mathbb{F}_q$ be a finite ring with $q=p^{m}$ and $p$ a prime number of the residue field $\mathbb{F}_{q}$.
Assume that the order of $\Theta$ is $2$ and $n$ is an even integer denoted by $n=2k$ such that $k$ is even.
Let $g(x)=\sum_{i=0}^{k-1} g_{i}x^{i}+x^{k}$ be a right divisor of $x^{n}-1$ with $g_{0}$ a unit in $R$.
Then the $\Theta-$cyclic code generated by $g(x)$ is Hermitian self-dual if and only if
\begin{equation}
\label{eq:bidon8}
gg^{\natural}=x^{n}-1.
\end{equation}
\end{prop}
\begin{proof}
Let $g(x)=g_{0}+g_{1}x+\ldots+g_{k-1}x^{k-1}+x^{k}$ be a right divisor of $x^{n}-1$.
Then by Definition \ref{def:bidon4}, the reciprocal of $g(x)$ is the polynomial
$g^{\ast}(x)=1+\Theta(g_{k-1})x+\dots+\Theta^{k-1}(g_{1})x^{k-1}+\Theta^{k}(g_{0})x^{k}$,
and the left monic skew reciprocal polynomial of $g(x)$ is
\[
g^{\natural}(x)=\frac{1}{\Theta^{k}(g_{0})}g^{\ast}(x)= \frac{1}{\Theta^{k}(g_{0})}\left(1+\Theta(g_{k-1})x+\dots+\Theta^{k-1}(g_{1})x^{k-1}+\Theta^{k}(g_{0})x^{k}\right).
\]
As the order of $\Theta$ is $2$ and from Proposition~\ref{prop:bidon8} Hermitian self-dual skew cyclic codes exist if $k$ is even, we have that
\begin{equation}
\label{eq:bidon 28}
g^{\natural}(x)=g_{0}^{-1}+g_{0}^{-1}g_{k-1}^{-1}x+\dots+g_{0}^{-1}g_{1}^{-1}x^{k-1}+x^{k}.
\end{equation}
Similarly, from Proposition~\ref{prop: bidon28} there exists a Hermitian self-dual skew cyclic code generated by a skew polynomial $g(x)$ if and only if
\begin{equation*}
\begin{array}{l}
(g_{0}+g_{1}x+\ldots+g_{k-1}x^{k-1}+x^{k})(\Theta^{-k-1}(g_{0}^{-1})+\Theta^{-k}(g_{0}^{-1}g_{k-1})x^{1}+\\
\hspace*{0.3in} \ldots+\Theta^{-2}(g_{0}^{-1}g_{1})x^{k-1}+x^{k})\\	
=(g_{0}+g_{1}x+\dots+g_{k-1}x^{k-1}+x^{k})(g_{0}^{-1}+g_{0}g_{k-1}^{-1}x+\dots+ g_{0}g_{1}^{-1}x^{k-1}+x^{k})\\
=x^{n}-1.
\end{array}
\end{equation*}
Then from (\ref{eq:bidon 28}), if $g_{0}$ is a unit in $R$ (\ref{eq:SDM3}) is equivalent to $gg^{\natural}=x^{n}-1$,
so from Lemma~\ref{lem:bidon 29} $gg^{\natural}=g^{\natural}g=x^{n}-1$.
Thus, (\ref{eq:SDM3}) equivalent to
\begin{equation}
\label{eq:bidon 30}
g^{\natural}g=x^{n}-1.
\end{equation}
\end{proof}
\begin{rem}
\label{rem:bidon 21}
From \cite[Section 2]{G}, the greatest common right divisor of $f$ and $g,$ denoted by $\mbox{gcrd}(f, g)$,
is the unique monic polynomial $w \in \mathbb{F}_{q}[x, \theta]$ of highest degree such that there exist
$u, v \in \mathbb{F}_{q}[x, \theta]$ with $f=uw$ and $g=vw$.
The least common right multiple of $f$ and $g$, denoted by $h=\mbox{lcrm}(f, g)$,
is the unique monic polynomial $w \in \mathbb{F}_{q}[x, \theta] $ of lowest degree such that there exist
$u, v \in \mathbb{F}_{q}[x, \theta] $ with $h= u f$ and $h=v g$.
\end{rem}

Before giving the construction algorithm for self-dual $\Theta-$cyclic codes we provide the following lemma.
\begin{lem}
\label{lem:bidon 10}
Let $f, g \in \Re$ such that $r=deg(f)$ and $s=deg(g)$ with $r\geq s$.
If the coefficients of $f, g$ are invertible, then the computation of $\mbox{gcrd}(f, g)$ and $\mbox{lcrm}(f, g)$ in $\Re$
is the same as in $\mathbb{F}_{q}[x, \theta]$.
\end{lem}
\begin{proof}
Consider $f(x), g(x) \in \Re $ such that $ f(x)=a_{0}+a_{1}x +\dots+a_{r}x^{r}$ and $g (x)=b_{0}+b_{1}x +\dots+b_{s}x^{s}$.
From \cite{SSP} we have that for $ s\leq r$,
$f(x)-a_{r}\Theta^{r-s}(b_{s}^{-1})x^{r-s}g(x)$ has degree less than that of $f(x)$.
Then iterating the above procedure by subtracting further left multiples of $g(x)$ from the result until the degree is less than the degree
of $g(x)$, we obtain skew polynomials $q(x)$ and $r(x)$ such that
\[
f(x)=g(x)q(x)+r(x) \mbox{ with } deg(r(x)) < deg(g(x)) \mbox{ or } r(x)=0.
\]
Note that $q(x)$ and $r(x)$ are unique.
Further, the division in $\Re$ is the same as in the ring $\mathbb{F}_{q}[x, \theta]$
and from Remark ~\ref{rem:bidon 21}, for $\mbox{gcrd}(f, g)$ and $\mbox{lcrm}(f, g)$ to be unique with $f, g\in \Re$,
the coefficients of $f$ and $g$ must be invertible.
Therefore, if the coefficients of $f, g$ are invertible then $\mbox{gcrd}(f, g)$ and $\mbox{lcrm}(f, g)$
can be obtained using the same procedure as for $\mathbb{F}_{q}[x, \theta]$.
\end{proof}

In the following algorithm, the computation of $g\in \Re$ such that $g_{0}$ is a unit in $R$ with the property $g^{\natural}g=x^{2k}-1$
(Proposition~\ref{prop: bidon 7}), is replaced with the computation of polynomials $g_{i}$ such that $g_{i}^{\natural}g_{i}=f_{i}^{p^{s}}(x^{2})$.
\begin{algorithm}[]
\caption{\textbf{Construction of self-dual $\Theta-$cyclic codes}}
\label{alg1}
\textbf{Input}: $n, R, \mathbb{F}_q$ of characteristic $p , \Theta \in Aut(R)$ of order $2$ and $\theta \in Aut(\mathbb{F}_q)$\\
\textbf{Output}: the set of all generator polynomials of self-dual $\Theta-$cyclic codes of length $n$ over $R$
\begin{enumerate}
\item compute $s$ and $t$ such that $k= p^{s}\times t$ with $k$ an integer and $\gcd(t,p)=1$
\item compute the factorization of $x^{n}-1=f_{1}^{p^{s}}(x^{2}) \dots f_{l}^{p^{s}}(x^{2})$,
 where the $f_{i}(x^{2})= f_{i}^{\natural}(x^{2})$ are pairwise coprime polynomials in
 $(\mathbb{F}_q)^{\theta}\left[x^{2}\right]=(\mathbb{F}_q)^{\theta}\left[y\right]$
\item for $i$ in $\{1 \ldots l\}$ do
\item compute the sets $\mathcal{G}_{i}=\{g_{i} \in \Re \mid g_{i}^{\natural}g_{i}=f_{i}^{p^{s}}(x^{2})\}$
\item end for
\item return $\mbox{lclm}(g^{\natural}_{1}, \dots, g^{\natural}_{l})\}$
\end{enumerate}
\end{algorithm}

\begin{rem}
\label{rem:bidon 30}
According to Lemma~\ref {lem:bidon 10}, $\mbox{lclm}(g^{\natural}_{1}, \dots, g^{\natural}_{l})$
in step $6$ of Algorithm \ref{alg1} can be computed using the extended Euclidean algorithm in \cite[Section 2]{G}.
Thus, this algorithm can be executed for $k$ odd for generator polynomials s$g=lclm(g^{\natural}_{1}, \dots, g^{\natural}_{l})$ with coefficient which are invertible.
\end{rem}

+\subsection{Algorithm Complexity}
In this subsection, the complexity of Algorithm \ref{alg1} is analyzed.
From Proposition~\ref{prop:bidon 9} we have that if $\gcd(t, p)=1$ with $n=2k=2\times t\times p^{s}$,
then $x^{n}-1$ has decomposition in $\Re$ given by
\begin{equation*}
 x^{n}-1=f_{1}^{p^{s}}(x^{2}) \dots f_{l}^{p^{s}}(x^{2}),
\end{equation*}
where the $f_{i}= f_{i}^{\natural}$ are pairwise coprime polynomials in $(\mathbb{F}_q)^{\theta}\subset \mathbb{F}_q[x, \theta]$.
Then the complexity of the skew factorization in $\Re$ is the complexity of the skew factorization in $\mathbb{F}_q[x, \theta]$
which is given in the following theorem.
\begin{thm}(\cite[Theorem 2.4.2]{xj})
\label{thm:bidon 9}
The skew factorization algorithm factors a skew polynomial of degree $n$ in $\mathbb{F}_q[x, \theta]$ with complexity
\[
\tilde{\mathcal{O}}(n2^{3}\log q+n \log^{2}q+n^{1+\epsilon}(\log q)^{1+o(1)}+F(n, (\mathbb{F}_q)^{\theta}),
\]
operations for $\epsilon > 0$.
Here, $F(n, \mathbb{F}_q)$ denotes the complexity
of the factorization of a (commutative) polynomial of degree $n$ over the finite field $\mathbb{F}_q$.
\end{thm}
From Theorem~\ref{thm:bidon 9} we have the following result.
\begin{thm}
\label{thm:bidon 10}
Let $R=\mathbb{F}_q+u\mathbb{F}_q$ be a finite ring with $q=p^{m}$ and $p$ odd prime number of the residue field $\mathbb{F}_{q}$.
Assume that the order of $\Theta$ is $2$ and $n$ is an even integer denoted by $n=2k$.
Let $g(x)=\sum_{i=0}^{k-1} g_{i}x^{i}+x^{k}$ be a right divisor of $x^{n}-1$.
Then the complexity of  Algorithm \ref{alg1} in
\[
\tilde{\mathcal{O}}(n2^{3}\log q+n \log^{2}q+n^{1+\epsilon}(\log q)^{1+o(1)}+F(n, (\mathbb{F}_q)^{\theta})+k^{2}2^{4}+k^{2}2+k2^{2}),
\]
 operations.
\end{thm}
\begin{proof}
We now consider the complexity of the steps of the algorithm.
In the following $n=2k$ and $k=t\times p^{s}$.
\begin{enumerate}
\item Computing the integers $s$ and $t$ takes $\tilde{\mathcal{O}}(1)$ operation in $R$.
\item As $\gcd(t, p)=1$, according to Proposition~\ref{prop:bidon 9} and Theorem~\ref{thm:bidon 9},
factoring the skew polynomial $x^{n}-1$ in $\Re$ takes
\[
\tilde{\mathcal{O}}(n2^{3}\log q+n \log^{2}q+n^{1+\epsilon}(\log q)^{1+o(1)}+F(n,(\mathbb{F}_q)^{\theta}),
\]
operations.
\item Before determining the complexity of obtaining $g_{i}^{\natural}g_{i}$, we outline the process.
Recall that for $g_{i}=\sum_{i=0}^{k} g_{i}x^{i} \in \Re$ we have $g_{i}^{\natural}=\frac{1}{\Theta^{k}(g_{0})} \sum_{i=0}^{k}\Theta^{i}(g_{k-i})x^{i}$ and
their product is
\begin{equation*}
\begin{array}{l}
g_{i}^{\natural}g_{i}=\frac{1}{\Theta^{k}(g_{0})}\sum_{i=0}^{k}\sum_{i=0}^{k} \Theta^{i}(g_{k-i}) x^{i} g_{i}x^{i}
=\frac{1}{\Theta^{k}(g_{0})}\sum_{i=0}^{k}\sum_{i=0}^{k} \Theta^{i}(g_{k-i}) \Theta^{i}(g_{i})x^{2i}.
\end{array}
\end{equation*}
Let $g_{k-i}=a_{k-i}+ub_{k-i} \in \Re$ so then $\Theta^{i}(g_{k-i})=\theta(a_{k-i})+u\theta(b_{k-i})$ where $a, b\in \mathbb{F}_q$ and $\theta\in \text{Aut}(\mathbb{F}_q)$.
From \cite[(3.1.1)]{xj}, $\Theta^{i}(g_{k-i})$ has complexity $\tilde{\mathcal{O}}(k 2^{2})$
because $\theta(a_{k-i})$ and $\theta(b_{k-i})$ can be computed in $\tilde{\mathcal{O}}(k 2^{2})$ operations
so $\Theta^{i}(g_{k-i}) \Theta^{i}(g_{i})$ can be computed in $\tilde{\mathcal{O}}(k^{2} 2^{4})$ operations
where $2$ is the order of $\Theta$ and $k$ is the degree of the skew polynomial $g$.
Once we have the coefficients, it remains to compute the product which is done in $O(k^{2})$ operations in $R$ so the
complexity of the skew product $g_{i}^{\natural}g_{i}$ is $\tilde{\mathcal{O}}(k^{2} 2^{4}+k^{2}2)$ operations.
\item The coefficients of $g_{i}$ are invertible so determining $\mbox{lclm}(g^{\natural}_{1}, \dots g^{\natural}_{r})$ over $R[x, \Theta]$
is the same as over $\mathbb{F}_{q}[x, \theta]$.
Thus from \cite{xj}, $\mbox{lclm}(g^{\natural}_{1}, \dots, g^{\natural}_{i})$ can be computed in $\tilde{\mathcal{O}}(k2^{2})$ operations
where $2$ is the order of $\Theta$ and $k$ is the degree of the skew polynomial $g$.
\end{enumerate}
Combining the complexities of the steps gives the desired result.
\end{proof}

\subsection{Computational Results}
To obtain examples using Algorithm \ref{alg1},
$x^{n}-1=f_{1}^{p^{s}}(x^{2}) \dots f_{l}^{p^{s}}(x^{2})$ must be factored where the $f_{i}$ are pairwise coprime polynomials in
$\mathbb{F}_{q}[x^{2}]=Z(\mathbb{F}_{q}[x, \theta])$ satisfying $f_{i}(x^{2})=f_{i}^{\natural}(x^{2})$, $i\in \{1, 2, \dots, l\}$.
Furthermore, Gr\"{o}bner basis computation is used to obtain the polynomials $f_{i}$ to compute the sets $G_{i}=\{g\in \Re; g^{\natural}g=f_{i} \}$.
Then Lemma~\ref{lem:bidon 10} can be used to compute the codes generated by
$g^{\natural}=\mbox{lclm}(g^{\natural}_{1}, g^{\natural}_{2},\dots, g^{\natural}_{l} ),$ for all $g_{i} \in G_{i}$.
The skew polynomials $g_{i} \in G_{i}$ are used to obtain the skew polynomial $g^{\natural}$ which generates a self-dual $\Theta-$cyclic code over $R[x, \Theta]$ of length $n$.

\begin{rem}
	A right factor of degree $n-k$ of $x^{n}-1$ generates a linear code with parameters $(n, k)$.
	If $\Theta$ is not the identity, then the factorization of polynomials in the skew polynomial ring $R[x, \Theta]$ is not unique,
	and in general $x^{n}-1$ has many more factors in $R[x, \Theta]$ compared to the factorization in the usual polynomial ring.
	The minimum distance of a code, denoted by $d_{min}$, can be calculated using MAGMA \cite{BO}.
% but for larger codes the time complexity is prohibitive.
%	Therefore, there is a limit to the size of the codes for which $d_{min}$ can be obtained.
	There can be many codes with the same parameters $(n, k, d_{min})$.
\end{rem}

\begin{expl}
Consider $\mathbb{F}_{4}+u \mathbb{F}_{4} $ where $\mathbb{F}_{4}=\mathbb{F}_{2}(a)$,
$\Theta$ the Frobenius automorphism $(\alpha \mapsto\alpha^{2})$ and
$\Re=(\mathbb{F}_{4}+u \mathbb{F}_{4})[x, \Theta]$, so then
\begin{equation*}
x^{6}-1=f_{1}(x^{2})f_{2}(x^{2}),
\end{equation*}
where
\begin{equation*}
\begin{array}{ccl}
f_{1}(x) &=& x^{2} + 1,\\
f_{2}(x) &=& x^{4}+ x^{2} + 1.
\end{array}.
\end{equation*}
The $f_{i}(x)$ are pairwise coprime polynomials in $\mathbb{F}_{2}[ x^{2}]=Z(\mathbb{F}_{2}[x, \theta])$
satisfying $f_{i}(x^{2})=f_{i}^{\natural}(x^{2})$, $i\in \{1, \dots, 2\}$.
Gr\"{o}bner basis computation is used to obtain the polynomials $f_{i}$ and then the sets $G_{1}=\{g\in \Re, g^{\natural}g=x^{2} + 1 \}$ and
$G_{2}=\{g\in \Re, g^{\natural}g=x^{4}+ x^{2} + 1 \}$ are obtained which gives
\begin{equation*}
\begin{array}{ccl}
G_{1}&=&\{x+u+1, x+1 \},\\
G_{2}&=&\{x^{2}+(u+1)x+1, x^{2}+x+1, x^{2}+\omega^{2}, x^{2}+\omega \}.
\end{array}.
\end{equation*}
The codes generated by $g=\mbox{lclm}(g_{1}^{\natural}, g_{2}^{\natural})$ for $g_{i} \in G_{i}$
can be obtained using Lemma ~\ref{lem:bidon 10}.
Further, from Remark~\ref{rem:bidon 30}, if $k$ is an odd integer,
then $g=\mbox{lclm}(g_{1}^{\natural}, g_{2}^{\natural})$ can be chosen as the generator polynomial
which has coefficients that are units in $R$.
Then the skew polynomials
\begin{equation*}
\begin{array}{ccl}
g_{1}& =&x+u+1,\\
g_{2} &=&x^{2}+(u+1)x+ 1,
\end{array}.
\end{equation*}
give the skew polynomial
\begin{equation*}
g = x^{3}+u+1
\end{equation*}
which generates a self-dual code over $\mathbb{F}_{4}+u \mathbb{F}_{4}$ of length $6$ and $d_{min}=4$.
The generator matrix of this code can be obtained from the generator polynomial using (\ref{equa:matrix}) as
\begin{equation}
G=
\begin{bmatrix}
1+u& 0& 0& 1& 0& 0\\
0& 1+u& 0 & 0& 1& 0\\
0& 0& 1+u& 0& 0& 1\\
\end{bmatrix}.
\end{equation}
Then using the Gray map (\ref{equa:map}) gives the matrix
	\begin{displaymath}
	\left[ \begin{array}{cccccccccccc}
	1& 0& 0& 0& 0& 0& 0& 1& 0& 0& 0& 0 \\
	0& 1& 0& 0& 0& 0& 1& 0& 0& 0& 0& 0 \\
	0& 0& 1& 0& 0& 0& 0& 0& 0& 1& 0& 0 \\
	0& 0& 0& 1& 0& 0& 0& 0& 1& 0& 0& 0 \\
	0& 0& 0& 0& 1& 0& 0& 0& 0& 0& 0& 1 \\
	0& 0& 0& 0& 0& 1& 0& 0& 0& 0& 1& 0
	\end{array}\right],
	\end{displaymath}
which generates a $[12, 6, 2]$ linear code over $GF(4)$.
\end{expl}

%\begin{rem}
%	\begin{enumerate}
%	\item[(i)] Magma cannot compute an author examples of Algorithm 1 because Gr$\ddot{o}$bner basis take a lot of time and memory.
%	\item[(ii)] When we compute $g=lcrm(g^{\natural}_{1}, \dots, g^{\natural}_{l})$ in the Algorithm \ref{alg1} with the property $gg^{\natural}=x^{2k}-1$ (Proposition~\ref{prop: bidon 7}), Magma don't give any result.
%\end{enumerate}
%  \end{rem}

\end{document}